\documentclass[a4paper]{jpconf}
\usepackage{bm,color}
\usepackage{graphicx}
\usepackage{amsmath}
\begin{document}
\title{Numerical Simulation of Spin-Chirality Switching in Multiferroics via Intense Electromagnon Excitations}
\author{Masahito Mochizuki$^{1,2}$ and Naoto Nagaosa$^{1,3}$}
\address{$^1$Department of Applied Physics, University of Tokyo,
Tokyo 113-8656, Japan}
\address{$^2$Multiferroics Project, ERATO (JST), Tokyo 113-8656, Japan}
\address{$^3$Cross-Correlated Materials Research Group (CMRG)
and Correlated Electron Research Group (CERG),
RIKEN-ASI, Wako 351-0198, Japan}

\begin{abstract}
Chirality, i.e., the right- and left-handedness of structure, is one of the key concepts in many fields of science including biology, chemistry and physics, and its manipulation is an issue of vital importance. The electron spins in solids can form chiral configurations. In perovskite manganites $R$MnO$_3$ ($R$=Tb, Dy,...etc), the Mn-spins form a cycloidal structure, which induces ferroelectric polarization ($\bm P$) through the relativistic spin-orbit interaction. This magnetism-induced ferroelectricity (multiferroics) and associated infrared-active spin waves (electromagnons) open a promising route to control the spins by purely electric means in a very short time. In this paper, we show theoretically with an accurate spin Hamiltonian for TbMnO$_3$ that a picosecond optical pulse can switch the spin chirality by intensely exciting the electromagnons with a terahertz frequency. 
\end{abstract}

\section{Introduction}
\label{Sec:Introduction}
Some structures are not invariant upon the mirror operation just like the right and left hands, and are called {\it chiral}. This concept is of fundamental importance in chemistry and biology. 
The selective chemical synthesis of chirality (called asymmetric synthesis) is crucial for reliable and safe productions of medicines and foods. Note that the chirality of molecules and crystals is hard to change once synthesized because it is determined by the chemical bonds. In physics, on the other hand, spins in magnets sometimes form a chiral order. The spin chirality is much more flexible since the spin is a quantum-mechanical and relativistic object, which offers an opportunity to control the chirality by external parameters. 

\begin{figure}[tdp]
\includegraphics[scale=1.0]{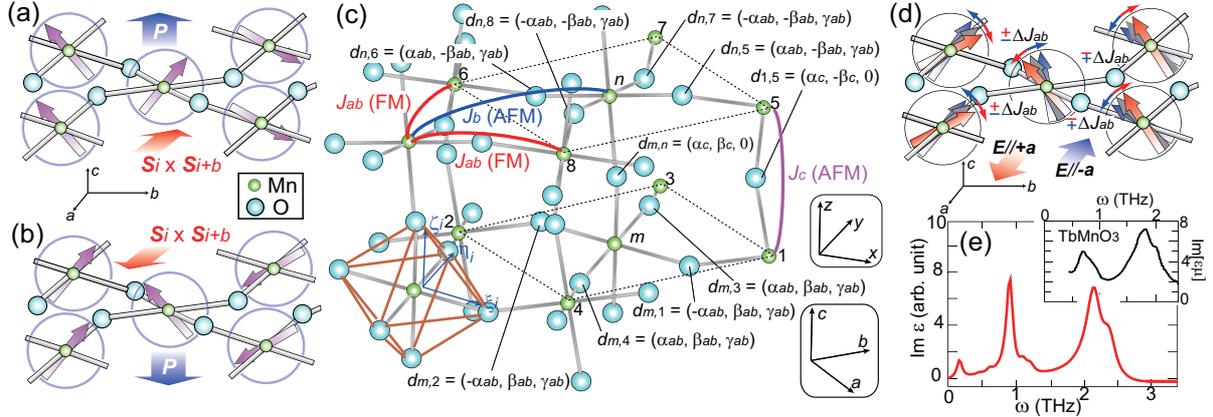}
\caption{(a) Spin configuration and ferroelectric polarization $\bm P$ in the clockwise $bc$-plane spin spiral and (b) those in the counterclockwise one. (c) Crystal structure, spin exchanges, Dzyaloshinskii-Moriya vectors $\bm d_{i,j}$, and local axes attached to the $i$th MnO$_6$ octahedron in $R$MnO$_3$. Here FM (AFM) denotes (anti)ferromagnetic exchange. (d) Modulation of the in-plane ferromagnetic exchanges under $\bm E$$\parallel$$\pm \bm a$, which induces slight spin rotations indicated by red (blue) arrows under $\bm E$$\parallel$$+\bm a$ ($\bm E$$\parallel$$-\bm a$). Upper (lower) signs in front of $\Delta J_{ab}$ ($>$0) correspond to the modulations under $\bm E$$\parallel$$+\bm a$ ($\bm E$$\parallel$$-\bm a$). (e) Calculated electromagnon optical spectrum. Inset shows the experimental spectrum for TbMnO$_3$ from Ref.~\cite{Takahashi08}.}
\label{Fig1}
\end{figure}
The representative laboratory for this spin chirality is the multiferroic transition-metal oxides~\cite{Review}. A model taking into account the spin-orbit interaction predicts the electric polarization $\bm p_{ij}$ produced by the chiral spin configuration~\cite{Katsura05,Mostovoy06} as 
\begin{equation}
\bm p_{ij} \propto \bm e_{ij} \times (\bm S_i \times \bm S_j),
\label{eq:PScrsS}
\end{equation}
with $\bm e_{ij}$ being the vector connecting $i$th and $j$th sites. Here the vector product $\bm S_i \times \bm S_j$ is the spin chirality, which characterizes a direction of the spin rotation as one goes from $i$ to $j$. This model explains the multiferroic behavior in $R$MnO$_3$ where the cycloidal Mn-spin order induces the ferroelectric polarization $\bm P$~\cite{Kimura03a}. For example, in TbMnO$_3$ [see Figs.~\ref{Fig1}(a) and (b)], the clockwise $bc$-plane spin spiral propagating along the $+\bm b$ direction induces $\bm P$$\parallel$$+\bm c$, while the counterclockwise one induces $\bm P$$\parallel$$-\bm c$~\cite{Kenzelmann05,Yamasaki07a}. For the $a$, $b$ and $c$ axes, we adopt the $Pbnm$ setting.

When the electric polarization is driven by the spin order, it is naturally expected that collective excitations of spins (magnons) have an optical activity~\cite{Smolenski82,Pimenov06a,Katsura07}. Indeed, strong optical absorptions were experimentally observed in $R$MnO$_3$ at terahertz frequencies, and they were ascribed to magnons activated by the electric-field component of light i.e., electromagnons~\cite{Pimenov06a,Katsura07,Pimenov08,Kida08,Kida08b,Takahashi08,Takahashi09,Kida09}. In the early stage, the corresponding magnon modes were interpreted as rotation of the spin-spiral plane accompanied by oscillation of $\bm p_{ij}$ in Eq.~(1)~\cite{Katsura07}. However, this interpretation contradicts to the experimental observation about the selection rule in terms of the light polarization~\cite{Kida08,Kida08b}. Afterwards, it turned out that the conventional magnetostriction mechanism where $\bm p_{ij}$ is given by
\begin{equation}
\bm p_{ij} =\bm \pi_{ij} (\bm S_i \cdot \bm S_j)
\label{eq:PSdotS}
\end{equation}
is relevant to the electromagnons in $R$MnO$_3$~\cite{Aguilar09,Mochizuki10}. Here the vector $\bm \pi_{ij}$ is nonzero because of the orthorhombic lattice distortion without inversion symmetry at the center of the Mn-O-Mn bond. The puzzling electromagnon optical spectrum with two specific peaks was successfully explained by this mechanism~\cite{Mochizuki10}. 

Under this circumstance, the photo-induced phenomena become a challenging issue for the next step. Since $\bm p_{ij}$ in Eq.~(2) does not require the spin-orbit interaction, its magnitude is much larger than that of $\bm p_{ij}$ in Eq.~(1), which enables us to realize the intense optical excitations of magnons. In fact, the real-time dynamics of the photo-excitation is usually a notoriously complicated problem since the electronic excitations are the genuine many-body problem of various degrees of freedom. Therefore, a reliable theoretical analysis has been almost impossible thus far. However, we are now ready to attack such phenomena in $R$MnO$_3$ theoretically. This is because we know that the optical pulse activates mostly the spin systems only via the magnetostriction given in Eq.~(2), which allows us to neglect electronic excitations at higher energies ($>$1.5 eV). In addition, we have an accurate spin Hamiltonian, which describes competitions among various phases in $R$MnO$_3$~\cite{Mochizuki09a}, so that the optical switching among them can be simulated in a reliable way. Recent rapid experimental progress in intense THz-pulse generation also gives a growing importance to the theoretical study on this subject~\cite{XieX06,Gaal07,Sell08}.

\section{Model and Method}
\label{Sec:ModelMethod}
To describe the Mn-spin system in TbMnO$_3$, we employ a classical Heisenberg model on a cubic lattice, which is given by
\begin{eqnarray}
\mathcal{H}&=&
\sum_{<i,j>} J_{ij} \bm S_i \cdot \bm S_j
+D \sum_{i} S_{\zeta i}^2
+E\sum_{i}(-1)^{i_x+i_y}(S_{\xi i}^2-S_{\eta i}^2) \nonumber \\
&+&\sum_{<i,j>}\bm d_{i,j}\cdot(\bm S_i \times \bm S_j)
-B_{\rm biq}\sum_{<i,j>}^{ab}(\bm S_i \cdot \bm S_j)^2.
\label{eq:HMLT}
\end{eqnarray}
Here $i_x$, $i_y$, $i_z$ represent the integer coordinates of the $i$th Mn ion with respect to the $x$, $y$ and $z$ axes. The first term describes the spin-exchange interactions as shown in Fig.~\ref{Fig1}(c). The second and the third terms stand for the single-ion anisotropy defined with local axes, $\xi_i$, $\eta_i$ and $\zeta_i$, attached to each MnO$_6$ octahedron tilted in the orthorhombic lattice~\cite{Mochizuki09a}. The fourth term denotes the DM interaction where the DM vectors $\bm d_{i,j}$ are expressed using five DM parameters, $\alpha_{ab}$, $\beta_{ab}$, $\gamma_{ab}$, $\alpha_c$ and $\beta_c$, as given in Ref.~\cite{Solovyev96}. The last term represents the biquadratic interaction originating from the spin-phonon coupling~\cite{Kaplan09}. The Monte-Carlo analysis of this model has successfully reproduced the rich phase diagram of $R$MnO$_3$~\cite{Mochizuki09a}. We perform calculations using following parameters: $J_{ab}$=$-$0.74, $J_b$=0.64, $J_c$=1.0, ($\alpha_{ab}$,$\beta_{ab}$,$\gamma_{ab}$)=(0.1, 0.1, 0.14), ($\alpha_c$,$\beta_c$)=(0.48, 0.1), $D$=0.2, $E$=0.25, and $B_{\rm biq}$=0.025, where the energy unit is meV. This parameter set gives the $bc$-plane transverse spin spiral propagating along the $b$ axis with a wave number $q_b$=0.3$\pi$ at low temperatures, which resembles the $bc$-plane spin spiral in TbMnO$_3$ ($q_b$=0.29$\pi$)~\cite{Kenzelmann05,Yamasaki07a}.

We trace dynamics of the Mn spins by numerically solving the Landau-Lifshitz-Gilbert equation using the fourth-order Runge-Kutta method~\cite{Mochizuki10b}. The equation is given by
\begin{equation}
\frac{\partial \bm S_i}{\partial t}=-\bm S_i \times \bm H^{\rm eff}_i
+ \frac{\alpha_{\rm G}}{S} \bm S_i \times \frac{\partial \bm S_i}{\partial t},
\label{eq:LLGEQ}
\end{equation} 
where $\alpha_{\rm G}$(=0.05) is the Gilbert-damping coefficient. We derive an effective magnetic field $\bm H^{\rm eff}_i$ acting on the spin $\bm S_i$ from the Hamiltonian $\mathcal{H}$ as $\bm H^{\rm eff}_i = - \partial \mathcal{H} / \partial \bm S_i$. Considering the strongly reduced Mn moment revealed in the neutron-scattering experiment~\cite{Arima06}, we set the norm of the spin vector $|\bm S_i|$=1.4. The system used for calculations is 20$\times$20$\times$6 in size with the periodic boundary condition.
For the magnetoelectric (ME) coupling, we consider $-\bm E \cdot \bm p_{ij}
=-(\bm E \cdot \bm \pi_{ij}) \bm S_i \cdot \bm S_j$
where $\bm p_{ij}$ is given in Eq.~(2)~\cite{Aguilar09,Mochizuki10}. This coupling effectively modulates the in-plane ferromagnetic exchanges from $J_{ab} \bm S_i \cdot \bm S_j$ to $(J_{ab}-\bm E \cdot \bm \pi_{ij}) \bm S_i \cdot \bm S_j$. More concretely, the applied $\bm E$$\parallel$$\pm \bm a$ modulates the spin exchanges as shown in Fig.~\ref{Fig1}(d). Here $|\pi_{ij}^a|$ is calculated to be 3.5$\times$10$^{-26}$ $\mu Cm$ from the lattice parameters\cite{Alonso00} and the observed $P$($\sim$5000$\mu C/m^2$) for HoMnO$_3$ with an up-up-down-down spin order~\cite{Ishiwata10}. This means that $E_a$=1 MV/cm induces the modulation $|\Delta J_{ab}|$=$|E_a \pi_{ij}^a|$=0.022 meV. Note that $\bm p_{ij}$ are perfectly canceled out in the equilibrium state, and never contribute to the static ferroelectricity.

The electromagnon spectrum is calculated as a response to the weak $\delta$-function pulse. We trace the time evolution of the net polarization $\bm P(t)=\sum_{<i,j>} \bm p_{ij}(t)$ after application of the pulse where $\bm p_{ij}(t)$ is given by Eq.~(2). Then we perform the Fourier transformation of $\bm P(t)$ to obtain Im $\varepsilon(\omega)$. The calculated spectrum is displayed in Fig.~\ref{Fig1}(e), which has two peaks at $\omega$=0.94 and 2.1 THz, and reproduces well the experimental spectrum of TbMnO$_3$~\cite{Takahashi08}.

Using this model, we theoretically demonstrate switching of the spin chirality by application of an intense optical pulse. Note first that there are two kinds of $bc$-plane spirals with different spin chiralities, i.e. clockwise and counterclockwise turns as one propagates along the +$\bm b$ direction. Their spin chiralities $\bm \chi$ point in the $-\bm a$ and $+\bm a$ directions ($\bm \chi$$\parallel$$-\bm a$ and $\bm \chi$$\parallel$$+\bm a$) so that they are referred to as $bc_{-}$ and $bc_{+}$, respectively. The chirality $\bm \chi$ is calculated as a sum of the local contributions $\bm \chi_{i,i+\hat x}=\bm S_i \times \bm S_{i+\hat x}$ and $\bm \chi_{i,i+\hat y}=\bm S_i \times \bm S_{i+\hat y}$ as 
$\bm \chi=\frac{1}{2N} \sum_{i} (\bm \chi_{i,i+\hat x} + \bm \chi_{i,i+\hat y})/S^2$. Note that $\bm \chi_{i,i+\hat z}$ is zero because of antiferromagnetic stacking in $z$-direction. The $ab$-plane spiral spin configuration is also possible in TbMnO$_3$ although it is slightly higher in energy than the $bc$-plane one without external fields. Again there are two kinds of $ab$-plane spirals, $ab_{-}$ and $ab_{+}$, which have $\bm \chi$$\parallel$$-\bm c$ and $\bm \chi$$\parallel$$+\bm c$, respectively.

\section{Results and Discussion}
\label{Sec:Results}
\begin{figure}[tdp]
\includegraphics[scale=1.0]{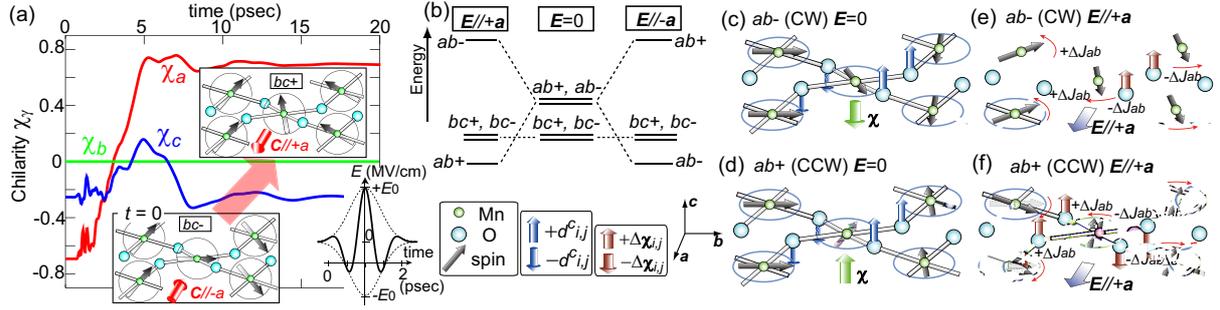}
\caption{(a) Time evolution of the spin chirality $\bm \chi$=($\chi_a$, $\chi_b$, $\chi_c$) after application of the pulse with $E_0$=11 MV/cm, which shows the chirality reversal. Inset shows spin states before/after applying the pulse, and time profile of the applied pulse $E_a(t)$. (b) Energy diagrams of the $ab_{\pm}$ and $bc_{\pm}$ states in the presence and absence of $\bm E$$\parallel$$\pm \bm a$. (c)[(d)] Direction of the net spin chirality $\bm \chi$ and $c$-axis components of the DM vectors $\pm d_{i,j}^c$ on the in-plane bonds for the clockwise [counterclockwise] $ab$-plane spin spiral, i.e. the $ab_{-}$ [$ab_{+}$] state when $\bm E$=0. (e) [(f)] Staggered modulation of the local spin chiralities $\bm \chi_{i,j} \pm \Delta \bm \chi_{i,j}$ under $\bm E$$\parallel$$+\bm a$ in the $ab_{-}$ [$ab_{+}$] state.}
\label{Fig2}
\end{figure}
Starting with $bc_{-}$ with $\bm \chi$$\parallel$$-\bm a$, we apply an intense pulse of electric field $\bm E(t)$=$[E_a(t), 0, 0]$ along the $a$ axis. Chirality switching by the application of an intense optical pulse has been theoretically proposed in Ref.~\cite{Mochizuki10b}. It has turned out that the switching processes show highly nonlinear behaviors with respect to the strength and shape of the pulse. Here we examine the pulse of
\begin{equation}
E_a(t)=E_0 \cos \omega t \exp[-\frac{t^2}{2\sigma^2}],
\end{equation}
which is different from the pulse with $\sin \omega t$ oscillation in the previous study~\cite{Mochizuki10b}. Here $\omega$ and $E_0$ are fixed at 2.1 THz and 11 MV/cm, while the full width of the half maximum of the Gaussian envelope is taken to be 0.5 psec [see inset of Fig.~\ref{Fig2}(a)]. In Fig.~\ref{Fig2}(a), we display calculated time evolution of the $a$-, $b$- and $c$-axis components of $\bm \chi$. We find a reversal of the spin chirality $\bm \chi$ from $bc_{-}$ ($\chi_a$$<$0) to $bc_{+}$ ($\chi_a$$>$0).

To understand this phenomenon, let us first consider energies of the four chirality states, i.e. $ab_{\pm}$ and $bc_{\pm}$. In Fig.~\ref{Fig2}(b), we summarize energy diagrams in the presence and absence of $\bm E$. When $\bm E$=0, the $ab_{+}$ and $ab_{-}$ states are degenerate, and are higher in energy than the ground-state $bc$-plane spirals. Application of $\bm E$$\parallel$$\pm \bm a$ lifts this degeneracy, and if it is strong enough, either $ab_{+}$ or $ab_{-}$ becomes the lowest-lying state depending on the sign of $\bm E$. Under the strong $\bm E$$\parallel$$+\bm a$ ($\bm E$$\parallel$$-\bm a$), the $ab_{+}$ ($ab_{-}$) state becomes the lowest in energy. In contrast, the energies of $bc_{\pm}$ are not affected by $\bm E$.

More concretely, when $\bm E$=0, the spins in both $ab_{+}$ and $ab_{-}$ states are rotating with nearly uniform turn angles, and all of the local spin chiralities $\bm \chi_{i,j}$=$\bm S_i \times \bm S_j$ have nearly the same amplitudes along $+\bm c$ and $-\bm c$ directions, respectively as indicated by green arrows in Figs.~\ref{Fig2}(c) and (d). The local chiralities $\bm \chi_{i,j}$ couple with the $c$-axis components of the DM vectors, $\pm d_{i,j}^c$, on the in-plane Mn($i$)-O-Mn($j$) bonds, whose magnitudes are all equal to $\gamma_{ab}$. Since the signs of $\pm d_{i,j}^c$ are staggered as shown by blue arrows, the total DM energy, $E_{\rm DM}=\sum_{i} (\bm d_{i,i+\hat x} \cdot \bm \chi_{i,i+\hat x} +\bm d_{i,i+\hat y} \cdot \bm \chi_{i,i+\hat y})$, cancels out for both $ab_{+}$ and $ab_{-}$ states, resulting in their degeneracy. When $\bm E$ is applied, the modulated ferromagnetic exchanges $J_{ab}-\Delta J_{ab}$ and $J_{ab}+\Delta J_{ab}$ cause slight spin rotations as indicated by thin red arrows in Figs.~\ref{Fig2}(e) and (f) to decrease and increase of the spin turn angles. Consequently the local spin chiralities become subject to a staggered modulation as $\bm \chi_{i,j} \pm \Delta \bm \chi_{i,j}$. The way of the modulation is different between $ab_{+}$ and $ab_{-}$, namely depending on the sign of $\bm \chi$. When $\bm E$$\parallel$$+\bm a$, the $+\Delta \bm \chi_{i,j}$ and $-\Delta \bm \chi_{i,j}$ are arranged antiparallel (parallel) to $\pm d_{i,j}^c$ in the $ab_{+}$ ($ab_{-}$) state, resulting in the decrease (increase) of DM energy. When this DM energy is large enough under the sufficiently strong $\bm E$$\parallel$$+\bm a$, the $ab_{+}$ state becomes the lowest in energy as shown in Fig.~\ref{Fig2}(b). In contrast, under the strong $\bm E$$\parallel$$-\bm a$, the $ab_{-}$ state becomes the lowest in energy. We note that energies of the $bc_{\pm}$ states never change even under $\bm E$ because the $a$-axis components of the DM vectors are alternately stacked, by which the DM energy cancels out despite the staggered modulation of the local chiralities, $\bm \chi_{i,j} \pm \Delta \bm \chi_{i,j}$ ($\parallel$$a$).

\begin{figure}[tdp]
\includegraphics[scale=1.0]{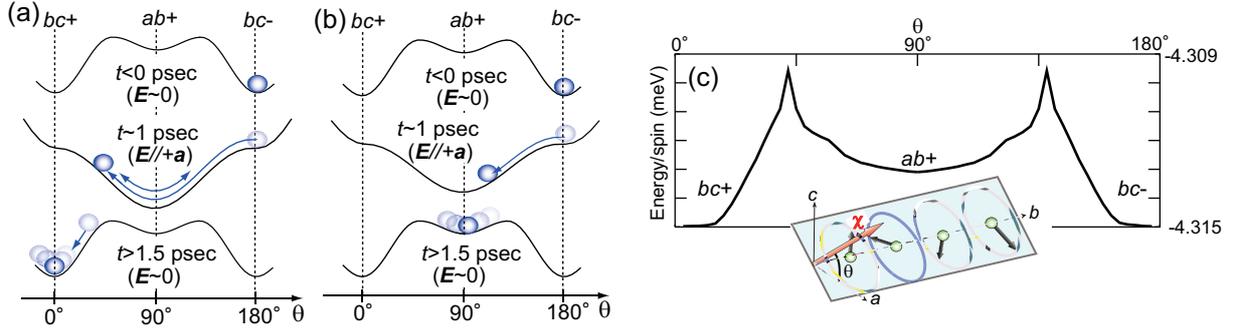}
\caption{(a) Schematic figure for time evolution of the energy structure in the $\theta$ space for the chirality-reversal process. (b) That for the chirality-flop process. (c) Calculated energy structure in the $\theta$ space at $\bm E$=0. Inset: Spin chirality $\bm \chi$ and the angle $\theta$ in the cycloidal spin state.}
\label{Fig3}
\end{figure}
Two key ingredients of this chirality switching are dynamical deformation of potential structure under the oscillating $\bm E$ via the above-discussed ME coupling and an inertial motion of the spin chirality. In Fig.~\ref{Fig3}(a), we depict a cartoon for time evolution of the potential structure in the $\theta$ space for the chirality-reversal process from $\theta$=180$^{\circ}$ ($bc_{-}$) to $\theta$=0$^{\circ}$ ($bc_{+}$). Here $\theta$ is the angle between the chirality $\bm \chi$ and the $a$ axis as defined in the inset of Fig.~\ref{Fig3}(c). When $E_a(t)$$>$0 as at $t$$\sim$1 psec, $\theta$=90$^{\circ}$ ($ab_{+}$) becomes a new energy minimum, so that the chirality $\bm \chi$ starts rotating towards this minimum. Importantly the chirality does not stop its rotation at $\theta$=90$^{\circ}$ immediately, but passes through that minimum or oscillates around it because of the inertial force. The DM interaction and the single-ion anisotropy originating from the spin-orbit coupling make the rotation of the spin-spiral plane massive, resulting in its inertial motion. 
When $E_a(t)$ becomes 0 again at $t>$1.5 psec, the system falls into the minimum at $\theta$=0$^{\circ}$ ($bc_{+}$) if the system is within the domain of metastability of $\theta$=0$^{\circ}$. The system can also be trapped in the local minimum at $\theta$=90$^{\circ}$ ($ab_{+}$), resulting in the 90$^{\circ}$ flop of the spin chirality [see Fig.~\ref{Fig3}(b)].

Here the equation of motion for the chirality oscillation around an energy minimum at $\theta$=$\theta_0$ can be written by introducing the effective mass $\tilde m$ as $\tilde m \ddot{\theta} + \tilde m \gamma \dot{\theta}=-\frac{\partial V(\theta-\theta_0)}{\partial \theta}$ where $\gamma$ is the damping rate determined by $\alpha_{\rm G}$. Note that the dimension of $\tilde m$ is not M but ML$^2$ since $\theta$ is a dimensionless variable in contrast to a variable appeared in the usual equation of motion. We calculate the $\theta$ dependence of energy by rotating the spin spiral plane around the $b$ axis, which corresponds to $V(\theta)$. We display calculated $V(\theta)$ for $\bm E$=0 in Fig.~\ref{Fig3}(c). Since the typical frequency of the oscillation is a few terahertz, the order of $\tilde m$ is estimated as 10$^{-18}$ sec$^2$eV.

This optical chirality switching requires a strong $\bm E$ ($|E_0|$$>$11 MV/cm) because it is necessary to lower the $ab$-spiral state energetically relative to the $bc$-spiral one. This threshold value can be reduced if we properly choose the materials: For example, the solid solutions Tb$_{1-x}$Gd$_x$MnO$_3$ locating near the boundary between the $ab$ and $bc$ spiral phases~\cite{Goto05} are promising candidates. 

\section{Summary}
\label{Sec:Summary}
In summary, we have theoretically demonstrated the optical switching of spin chirality by intensely exciting the electromagnons in multiferroic TbMnO$_3$. We have revealed that the oscillating electric-field component of light activates collective rotations of the spin-spiral plane with terahertz frequencies via the ME coupling, and their inertial motions result in the chirality reversal. The collective spin motion is crucial to this phenomenon, which supplies energy to cross the potential barriers. The mechanism here proposed is generic, and can be expected also in other spin-spiral-based multiferroics.

The authors are grateful to N. Kida, Y. Tokura, N. Furukawa, R. Shimano and I. Kezsmarki for fruitful discussions. This work was supported by Grant-in-Aid (Grants No. 22740214, No. 21244053, No. 17105002, No. 19048015, and No. 19048008) and G-COE Program ``Physical Sciences Frontier" from MEXT Japan, and Funding Program for World-Leading Innovative R$\&$D on Science and Technology (FIRST Program) from JSPS.


\end{document}